\documentclass[12pt]{article}
\usepackage{cite,graphicx}

\def\beq{\begin{equation}}
\def\eeq{\end{equation}}
\def\bea{\begin{eqnarray}}
\def\eea{\end{eqnarray}}
\newcommand{\gsim}{\lower.7ex\hbox{$\;\stackrel{\textstyle>}{\sim}\;$}}
\newcommand{\lsim}{\lower.7ex\hbox{$\;\stackrel{\textstyle<}{\sim}\;$}}

\begin{document}

\thispagestyle{empty}
\vspace*{.5cm}
\noindent
DESY 04-109 \hspace*{\fill} July 2, 2004\\
\vspace*{1.6cm}

\begin{center}
{\Large\bf Gauge Unification in Extra Dimensions:\\[0.3cm]
Power Corrections vs. Higher-Dimension Operators}
\\[2.0cm]
{\large A. Hebecker and A. Westphal}\\[.5cm]
{\it Deutsches Elektronen-Synchrotron, Notkestrasse 85, D-22603 Hamburg,
Germany}
\\[1cm]

{\bf Abstract}\end{center}
\noindent
Power-like loop corrections to gauge couplings are a generic feature of
higher-dimensional field theories. In supersymmetric grand unified theories 
in $d=5$ dimensions, such corrections arise only in the presence of a vacuum 
expectation value of the adjoint scalar of the gauge multiplet. We show 
that, using the analysis of the exact quantum effective action by 
Intriligator, Morrison and Seiberg, these power corrections can be 
understood as the effect of higher-dimension operators. Such operators, 
both classical and quantum, are highly constrained by gauge symmetry and 
supersymmetry. As a result, even non-perturbatively large 
contributions to gauge coupling unification can be unambiguously 
determined within 5d low-energy effective field theory. Since no massive 
hypermultiplet matter exists in 6 dimensions, the predictivity 
is further enhanced by embedding the 5d model in a 6d gauge theory 
relevant at smaller distances. Thus, large and quantitatively controlled 
power-law contributions to gauge couplings arise naturally and can, in the 
most extreme case, lead to calculable TeV-scale power law unification. 
We identify a simple 5d SU(5) model with one massless ${\bf 10}$ in the 
bulk where the power-law effect is exactly MSSM-like.

\newpage

\section{Introduction}

The elegance with which the Standard Model (SM) gauge and matter fields fit 
into the groups SU(5)~\cite{gg} or SO(10)~\cite{gfm} and their smallest 
representations forms the main piece of evidence for the idea of Grand 
Unified Theories or GUTs (also~\cite{ps}). In the conventional view, this 
is further supported by quantitative gauge coupling unification at 
$\sim 10^{16}$ GeV due to logarithmic running in 4d supersymmetric (SUSY) 
field theory~\cite{gqw,drw}. However, given the excellent theoretical and 
phenomenological motivation of extra dimensions, especially in the context
of string-theoretic~\cite{chsw} or field-theoretic~\cite{orb5a,orb5b,orb6,wy} 
unified models, it is important to understand novel features introduced by 
loop-effects in higher dimensions.

It is by now well-known that power-like loop corrections in higher 
dimensions can affect gauge unification in a dramatic way~\cite{ddg} 
(see also~\cite{tv}). However, their quantitative analysis requires 
the detailed knowledge of the high-scale unified model~\cite{cprt,obj}. 
Related issues have also been discussed in the context of dimensional
deconstruction (see, e.g.,~\cite{dec}).

As described in detail in~\cite{hw} (cf. also~\cite{ops}), the dominant 
power-law effects are calculable within effective field theory (i.e., are 
independent of the UV completion) if the 5- or 6-dimensional unified gauge 
theory is broken by the vacuum expectation value (VEV) of a bulk scalar 
field. The relevant finite loop correction, linear in the mass or momentum 
scale in $d\!=\!5$, was also obtained in the context of AdS models~\cite{gr1, 
cct,gr}. In particular, the linear correction to gauge unification is 
implicitly contained in~\cite{cct} (cf.~also~\cite{ads1}). 

Specifically, the loop corrections to the differences of the inverse SM 
gauge couplings $\alpha_4^{-1}$ in 4 dimensions, evaluated directly below 
the compactification scale $M_c\sim 1/R$, take the form~\cite{hw}
\beq
\Delta\alpha_4^{-1}(M_c)\sim\left\{\begin{array}{ll}\,\,\,|\Phi| R & 
\textrm{, $d=5$}\\ (\,|\Phi|R\,)^2\cdot\ln\left(\Lambda/|\Phi|\right) & 
\textrm{, $d=6$\,.}\end{array} \right. 
\label{corr} 
\eeq 
Here $\Phi$ is the relevant GUT-breaking VEV and $\Lambda$ is the UV cutoff 
scale (implying that, in $d=6$, only the leading-log term is predicted).
Calculability holds in the same sense as for threshold corrections
of 4d GUTs~\cite{wei1,thresh,fra}, i.e., they are uniquely specified by 
group theory once the field content and relevant mass thresholds of the 
model are given.

A potential problem comes from higher-dimension operators such as $(1/ 
\Lambda)^n\,$tr$[\Phi^n\,F^2]$. If $\Phi$ develops a VEV, they contribute 
to gauge coupling differences at the tree level and compete with the 
calculable loop effects~\cite{cct,gr}. The most dangerous such operators may 
be forbidden or restricted by symmetries~\cite{hw}. Nevertheless, their 
presence limits the quantitative control over power-law corrections to the 
region $|\Phi|/\Lambda\ll 1$ in the non-SUSY case. In particular, consider 
a 5-dimensional model where $\Lambda\sim g_5^{-2}$ sets the fundamental 
scale of the bulk theory. Then the requirement $|\Phi|\ll\Lambda$ implies 
that the corrections of Eq.~(\ref{corr}) can not change the tree-level 
relation $(4\pi\alpha_4)^{-1}=g_4^{-2}=2\pi Rg_5^{-2}$ at the 100\% level.

The present paper demonstrates that, in the supersymmetric case, the 
analysis of the exact quantum effective action by Intriligator, Morrison 
and Seiberg~\cite{ims} drastically improves the situation outlined above. 
The statements of~\cite{hw} concerning the predictive power and validity 
range of power-law calculations are strengthened very significantly. 
Here the crucial point is that the quantum effective action at the 
two-derivative level is completely known. Technically, this follows from 
the SUSY-based restrictions on 
higher-dimension operators, including the absence of two-derivative 
operators of mass dimension 6 and higher in the super Yang-Mills (SYM) 
lagrangian. (Higher-derivative operators 
can be present but do not affect low-energy gauge couplings.) In particular, 
low-energy 4d gauge couplings may receive 100\% corrections from 
higher-dimensional power-law effects which are nevertheless controlled 
within effective field theory. In the most extreme case, this allows 
scenarios with quantitative TeV-scale power-law unification.

The paper is organized as follows. In Sect.~\ref{prep} we recall the quantum 
exact prepotential of 5d SYM theory discussed in detail 
in~\cite{ims} (for earlier results see \cite{gst} and, in particular,
\cite{aft}). For our purposes, it is essential that the only classical 
operators are the (SUSY versions of) gauge kinetic and Chern-Simons (CS) 
terms and that quantum corrections, which arise only at the 1-loop-level, 
are known explicitly. 

In Sect.~\ref{pl} it is shown how the by now familiar power-corrections 
to gauge unification (sometimes referred to as `power-law running') arise 
in the above framework. They correspond to higher-dimension operators 
which are, in general, non-analytic in the symmetry-breaking VEV. Corrections 
induced by a bulk hypermultiplet become analytic (in fact, identical to a 
classical CS term) if the hypermultiplet mass is sufficiently 
large. However, the tuning of finite hypermultiplet masses comparable 
to the bulk VEV allows the realization of almost any desired power law
effect. This ambiguity is avoided if the 5d model arises as the low-energy 
limit of a 6d construction because of the absence of massive hypermultiplets 
in 6 dimensions.

Section~\ref{5d} introduces a realistic SU(5) model on $S^1/Z_2$, where the 
GUT group is broken to the SM group at both boundaries. The bulk field 
$\Phi$ breaking the symmetry in 5d is the adjoint scalar of the 5d vector 
multiplet. Its VEV, which is stabilized, e.g., by boundary Fayet-Iliopoulos 
(FI) terms, induces large power-law corrections to gauge unification. At 
the same time, this VEV gives masses to the $A_5$ zero modes of $X,Y$ gauge 
bosons which would otherwise plague an SU(5) orbifold GUT on $S^1/Z_2$. 
Power-law corrections to gauge unification are given in terms of the 
$\Phi$-VEV, the bulk hypermultiplet masses, and the bulk CS term, the 
latter being fixed by brane anomaly cancellation. Intriguingly, a bulk 
field content of just the gauge multiplet and a massless ${\bf 10}$ of 
SU(5) induces a power-law effect that is identical to the logarithmic 
running within the Minimal Supersymmetric Standard Model (MSSM). The extreme 
lightness of one of the three SM generations emerges naturally in this 
context. 

In Sect.~\ref{6d}, power-like loop corrections in 6-dimensional unified 
models are considered~\cite{gng,ghi} (for related earlier string theory 
results, especially including Wilson lines, see, e.g.,~\cite{str} 
and~\cite{wil,bac}). If the geometry is equivalent to $T^2/Z_2$ and 
$R_5\gg R_6$, the phenomenology of such models becomes very similar
to the 5d case. In particular, the finite 5d version of Eq.~(\ref{corr}) 
rather than the logarithmically UV-sensitive 6d version applies. The role 
of the $\Phi$-VEV is taken over by an $A_6$ Wilson line wrapping the 
cylinder-like central part of the compact space. The value of this Wilson 
line is fixed, e.g., by the orbifold breaking of the gauge symmetry at the 
4d fixed points. As far as power-law corrections are concerned, such 
effectively 5d scenarios arising from the compactification of 6d theories
are more predictive than pure 5d models because of the absence of massive 
gauged hypermultiplets and 6d anomaly constraints on massless bulk matter. 

The summary and our conclusions can be found in Sect.~\ref{con} and 
some technical details are given in the Appendix.

\section{Prepotential of the 5d SYM theory}\label{prep}
In this Section, we collect the relevant results of~\cite{ims,gst,aft} and set
up the notation used in the rest of the paper. Consider a 5d SYM theory 
with massive gauged hypermultiplet matter. In addition to the vector field 
and gaugino, the 5d vector multiplet contains an adjoint scalar field $\Phi$. 
The theory is conveniently described as a 4d ${\cal N}\!=\!2$ SYM theory 
depending on the extra parameter $x^5$. Its low-energy effective action is 
thus completely characterized by its holomorphic prepotential 
${\cal F}(\Sigma)$~\cite{seib} (see also~\cite{sw}). The scalar component 
of the chiral superfield $\Sigma$ is $\Phi+iA_5$, where we use the 
conventions of~\cite{heb} (see also~\cite{agw}) but interchange the names 
$\Sigma$ and $\Phi$ to facilitate comparison with~\cite{ims}. Given the 
prepotential, the lagrangian of a 4d ${\cal N}\!=\!2$ SYM theory can be 
written in conventional ${\cal N}\!=\!1$ superfield notation as 
\beq
{\cal L}=\frac{1}{2}\left\{\int d^4\theta\,\,\frac{\partial{\cal F}(\Sigma)}
{\partial\Sigma^a}\,\,\left(\bar{\Sigma}e^{2V}\right)^a\,+\,\int d^2\theta
\,\,\frac{\partial^2{\cal F}(\Sigma)}{\partial\Sigma^a\,\partial\Sigma^b}
\,\,W^aW^b\right\}\,+\,\mbox{h.c.}\label{pre}
\eeq
Here $\Sigma=\Sigma^aT_a$ and the generators of the gauge group $G$ are 
normalized by $2\,$tr$T_aT_b=\delta_{ab}$ (traces are taken in the 
fundamental representation unless otherwise specified).

Under the constraints of SUSY and 5d Lorentz invariance, the 4d lagrangian of 
Eq.~(\ref{pre}) extends in a unique way to a 5d lagrangian. However, 5d 
gauge invariance now constrains the prepotential to be at most cubic in 
$\Sigma$. In our context, this is crucial since it ensures the absence of 
higher-dimension operators beyond the CS term (see 
Appendix~A for more details). Following~\cite{ims}, we can also 
write the prepotential as a function of $\Phi$. Requiring the prepotential 
to be analytic, the most general form is now
\beq
{\cal F}(\Phi)=\frac{1}{2g_{5,cl.}^2}\,\mbox{tr}\,\Phi^2+\frac{c_{cl.}}
{48\pi^2}\,\mbox{tr}\,\Phi^3\,.\label{calf}
\eeq
The coefficients of these two terms determine the coefficients of the 
classical $F^2$ term and of the classical CS term, all other 
terms in the component lagrangian then being fixed by supersymmetry. (The
normalization is chosen such that, in the absence of charged matter, 
$c_{cl.}$ is integer due to the boundary anomaly constraint. This will become 
evident below.) In the present context of gauge coupling unification, it 
is crucial that the SUSY CS term includes an operator $\sim\Phi 
F^2$, which clearly has the potential of affecting low-energy gauge 
couplings if $\Phi$ develops a VEV. Thus, the most important two terms of 
the component lagrangian derived from Eq.~(\ref{calf}) are 
\beq
{\cal L}\supset -\frac{1}{2g_{5,cl.}^2}\,\mbox{tr}\,F^2-\frac{c_{cl.}}
{16\pi^2}\,\mbox{tr}\,\Phi F^2\,.\label{lag}
\eeq

The field $\Phi$ has a flat potential and one can consider the low-energy 
effective field theory in the presence of a $\Phi$-VEV. It will become 
clear from the discussion in Sect.~\ref{6d} that such a $\Phi$-VEV (rather 
than just hypermultiplet VEVs) is necessary in order for loop corrections 
to gauge unification to arise. Without loss of 
generality, we write $\Phi=\phi^iH_i$, where $H_i$ are the Cartan generators 
of the gauge group $G$ and $i\in\{1,\cdots,r=\,$rank$(G)\}$. We choose the 
$H_i$ to be the first $r$ elements of the set of generators $T_a$. Since a 
generic VEV breaks $G$ to U(1)$^r$, the relevant quantity is the 
prepotential of this abelian gauge theory. Including quantum corrections 
induced by the vector and hypermultiplets and choosing counterterms such that
$g_{5,cl.}$ and $c_{cl.}$ remain unchanged, it reads~\cite{ims,wit}
\beq
{\cal F}(\Phi)=\frac{1}{4g_{5,cl.}^2}\delta_{ij}\phi^i\phi^j+\frac{c_{cl.}}
{48\pi^2}d_{ijk}\phi^i\phi^j\phi^k+\frac{1}{96\pi^2}\left(\sum_\alpha|
\alpha_i\phi^i|^3-\sum_f\sum_\lambda|\lambda_i\phi^i+m_f|^3\right)\,.
\label{lc}
\eeq
Given the definition
\beq
d_{abc}=\frac{1}{2}\,\mbox{tr}\,T_a\{T_b,T_c\}\,,
\eeq
it is clear that the first two terms of Eq.~(\ref{lc}) are simply a
restriction of Eq.~(\ref{calf}) to the U(1)$^r$ subgroup. The remaining terms 
are the 1-loop-effects resulting from integrating out the heavy vector
multiplets (corresponding to the broken directions of $G$) and the 
hypermultiplets with masses $m_f$ labelled by their `flavour' $f$. The other 
sums run over the roots $\alpha$ of Lie$(G)$ and the weights $\lambda$ of 
the relevant matter representations\footnote{If several representation 
vectors have the same weight vector $\lambda$, this weight vector 
contributes with the appropriate multiplicity.}. Our notation implies that
\beq
[H_i,E_\alpha]=\alpha_i E_\alpha\qquad\mbox{and}\qquad H_i|\lambda\rangle=
\lambda_i|\lambda\rangle\,,
\eeq
where $E_\alpha$ is the Lie algebra element (root) corresponding to the 
root vector $\alpha$ and $|\lambda\rangle$ is a representation vector
with weight vector $\lambda$ (see, e.g.,~\cite{sla}). It is important that
Eq.~(\ref{lc}) is interpreted as defining a locally {\it holomorphic} 
prepotential, i.e., the modulus-signs merely determine whether a given 
cubic term is to be multiplied by $+1$ or $-1$ in a given region of the 
multi-dimensional space parameterized by $\phi^i$. Note also that the 
coefficient of the last term in Eq.~(\ref{lc}) differs from Ref.~\cite{ims} 
due to our different normalization of $c_{cl.}$. 

For our purposes, it is essential that Eq.~(\ref{lc}) specifies the 
complete low-energy effective action -- no higher-loop contributions 
arise and no other classical terms are allowed at the two-derivative 
level.

As done before in Eq.~(\ref{lag}) for the classical non-abelian theory,
we now give the gauge-kinetic term of the component lagrangian for 
each of the surviving U(1) factors. For the U(1) group generated by 
$H_i$ the relevant piece of the component lagrangian reads
\beq
{\cal L}_i\supset -\frac{1}{4}F_i^2\left\{\frac{1}{g_{5,cl.}^2}+
\frac{c_{cl.}}{4\pi^2}d_{iij}\phi^j+\frac{1}{8\pi^2}\left(\sum_\alpha
\alpha_i^2|\alpha_j\phi^j|-\sum_f\sum_\lambda\lambda_i^2|\lambda_j\phi^j+
m_f|\right)\right\}\,.\label{gc}
\eeq

\section{Power-law corrections from higher-dimension\\ operators}\label{pl}
It is now very easy to see that the above result corresponds precisely to
the power-like loop corrections to gauge unification considered recently 
by many authors following the proposal of~\cite{ddg}. The one-loop 
correction to a U(1) gauge coupling induced by massive particles is a 
standard result in quantum field theory. It is the basic building block 
of GUT threshold calculations~\cite{wei1,thresh} (see also~\cite{fra}). 
In particular, the dimensionally regularized result of~\cite{wei1} lends 
itself to an immediate implementation in the 5d situation. Combining the 
effects two complex scalars and a Dirac fermion, as appropriate for a 
massive hypermultiplet with mass $m$ and charge $q$, the correction reads
\beq
\delta\left(\frac{1}{g_5^2}\right)=-\frac{q^2}{8\pi^2}\,m\,.\label{lo}
\eeq
This dimensionally regularized result hides a mass-independent, linearly 
divergent piece. However, this piece is irrelevant in the present context 
since it is universal with respect to the different U(1) subgroups emerging 
from a spontaneously broken simple group $G$. It gives rise to a 
renormalization of the original non-abelian gauge coupling. 

In the context of the previous section, the hypermultiplet component 
corresponding to the weight $\lambda$ has mass $|\lambda_j\phi^j+m_f|$ and,
with respect to the U(1) subgroup generated by $H_i$, charge $\lambda_i$.
Thus, Eq.~(\ref{lo}) precisely reproduces the matter contribution in 
Eq.~(\ref{gc}). Furthermore, the correction from massive vector 
multiplets in Eq.~(\ref{gc}) is also unambiguously determined since, 
apart from the different charges and masses, its contribution must be
equal and opposite in sign compared to the hypermultiplet. This is clear 
since a vector- and hypermultiplet with the same mass and charge combine to
a 4d ${\cal N}\!=\!4$ multiplet and therefore induce no gauge coupling 
correction. 

Of course, we are not really interested in the breaking of a gauge group
$G$ of rank $r$ to U(1)$^r$ but rather in its breaking to a set of simple 
subgroups and U(1) factors, for which we use the common notation $G_i$. 
For example, $G_i$ with $i=1,2,3$ may be the three gauge groups of the SM. 
The relevant gauge coupling corrections can be immediately read off from 
Eq.~(\ref{gc}) by choosing an appropriate $\Phi$-VEV, i.e., appropriately 
degenerate $\phi^i$. It is useful to present the corresponding result in a 
different form, using traces of representation generators. In this form, 
the correction to the low-energy gauge coupling of the subgroup $G_i$ reads
\beq
\delta\left(\frac{1}{g_{5,i}^2}\right)=\frac{c_{cl.}}{4\pi^2}\,\mbox{tr}\,
H_i^2\Phi\,+\frac{1}{8\pi^2}\left(\sum_{r_i(a)}T_{r_i(a)}M_{r_i(a)}-\sum_f
\sum_{r_i(f)}T_{r_i(f)}M_{r_i(f)}\right)\,.\label{gct}
\eeq
Here $H_i$ is one of the Cartan generators of $G$ that fall into $G_i$. 
The $G_i\,$-representations emerging from the adjoint of $G$ and from the 
representation of the hypermultiplet $f$ are labelled by $r_i(a)$ and 
$r_i(f)$ respectively. As usual, $T_{r_i}$ is defined by tr$_{r_i}[T_aT_b]=
\delta_{ab}T_{r_i}$, with the trace taken in the representation $r_i$. 
Furthermore, $M_{r_i(a)}$ and $M_{r_i(f)}$ denote the masses of the vector 
multiplet in the representation $r_i(a)$ and the hypermultiplet in the 
representation $r_i(f)$ respectively. Given these definitions and the 
relations
\beq
\sum_{\alpha\,\in\,r_i(a)}\alpha_i^2=\mbox{tr}_{r_i(a)}H_i^2=T_{r_i(a)}
\qquad,\qquad \sum_{\lambda\,\in\,r_i(f)}\lambda_i^2=\mbox{tr}_{r_i(f)}
H_i^2=T_{r_i(f)}\,,
\eeq
the derivation of Eq.~(\ref{gct}) from Eq.~(\ref{gc}) is straightforward. 

So far, we have just recovered the 5d threshold formulae of~\cite{hw}, 
based on the 4d results of~\cite{wei1,thresh}, in the prepotential 
language of~\cite{ims}, which is based on the anomaly calculation 
of~\cite{wit}. However, this deeper conceptual understanding of
power-like threshold corrections is crucial for their phenomenological 
applicability. The main point here is that the above prepotential formulae 
are quantum exact, which implies that the by now familiar 1-loop power-law
contributions to gauge unification are not subject to further corrections. 
More specifically, while higher-loop contributions are absent because of ${
\cal N}\!=\!2$ SUSY, the only competing tree-level higher-dimension operator 
is the SUSY CS term, corresponding to the first term on the r.h. side 
of Eq.~(\ref{gct}). (Note also that the holomorphic gauge couplings discussed 
here coincide with the canonical gauge couplings in ${\cal N}\!=\!2$ 
SUSY~\cite{hol}.) Moreover, in the phenomenologically relevant case of 
a compactification on an interval, the CS term induces anomalies
at the boundaries~\cite{ch} (see~\cite{ss} for a recent review). These 
induced anomalies must precisely cancel possible boundary anomalies coming 
from gauged bulk or brane fields. Thus, the value of the coefficient 
$c_{cl.}$ is completely determined by the field content of the model. 
This will be worked out in more detail in Sect.~\ref{5d}. 

The high predictivity of this scenario relies on the uniqueness of the 
tree-level dimension-5 operator, i.e., the SUSY CS term. This 
uniqueness is clearly based on the analyticity of the prepotential as a 
function of $\Phi$ and the uniqueness of the third-order symmetric 
invariant tensor $d_{abc}$~\cite{rsv} (in fact, such an invariant exists 
only for SU($N$) groups). Furthermore, it is also clear that the quantum 
corrected prepotential is not globally analytic (it is only analytic 
away from points where certain charged particle masses vanish). This allows 
for the distinct group-theoretical structures appearing in the quantum part
of Eq.~(\ref{lc}). However, for a given $\Phi$-VEV, any of the 
hypermultiplet contributions becomes analytic in the limit $|m_f|\to\infty$. 
In fact, because of the relations 
\beq
2\sum_\lambda \lambda_i\lambda_j \sim \delta_{ij} \qquad\mbox{and}\qquad
\sum_\lambda \lambda_i\lambda_j\lambda_k \sim d_{ijk}\,,\label{pro}
\eeq
it simply corrects the already existing tree-level operators $\sim 
\delta_{ij}$ and $\sim d_{ijk}$. (For the fundamental representation the 
proportionalities in Eq.~(\ref{pro}) become equalities.) In this sense, 
heavy matter effectively decouples from gauge unification corrections, 
i.e., its only trace is a contribution to the CS term which, 
however, is anyway fixed by low-energy anomaly constraints. 

Finally, one may consider the following somewhat exotic possibility. If
a certain hypermultiplet is in a large representation, then $\lambda_i
\phi^i$ can balance even a very large $m_f$ and a non-analytic contribution 
to Eq.~(\ref{lc}) may result. However, this does not contradict the above 
claim of effective decoupling since, given the spread of the values of 
$\lambda_i$ in a large representation, many relatively light states (with a 
mass comparable to $|\Phi|$) will 
automatically also be present. Thus, the presence of a large-representation
hypermultiplet will be known to the low-energy effective field theorist
even if its mass is very large.

We can now conclude that in a 5d SYM theory with hypermultiplet matter which
is broken by the VEV of the scalar adjoint $\Phi$, power-law corrections to
gauge unification are calculable in low-energy effective field theory.

\section{5d GUT phenomenology}\label{5d}
\subsection{Basic structure}
The simplest scenario in which the above power-law corrections to 5d
low-energy gauge couplings become relevant for a realistic GUT model
is that of a field-theoretic $S^1/Z_2$ orbifold~\cite{kaw} (see also
the slightly different later models of~\cite{orb5a,orb5b}). Specifically, 
consider a 5d SYM theory with gauge group $G$ and hypermultiplet matter 
compactified on an $S^1$ parameterized by $x^5\in[0,2\pi R)$ and restrict 
the field space by requiring invariance under the reflection $x^5\to -x^5$. 
If the space-time action of this $Z_2$ is accompanied by an inner 
automorphism of $G$ (characterized by an element $P\in G$ with $P^2=1$) 
acting in field-space, the gauge group is broken at both boundaries. In 
general, the surviving subgroup contains a U(1) factor which contains 
$P$, i.e., $G\supset G'\times$U(1). We now assume that boundary interactions 
stabilize a VEV of the adjoint scalar $\Phi$ which points in the direction 
of the U(1) generator~(cf.~\cite{mar}). This breaks the gauge group in the 
bulk in the same way as the orbifolding does at the two boundaries.

The $\Phi$-VEV can, for example, be stabilized by introducing a 
FI term within the U(1) subgroup surviving at each 
boundary. This term is, in general, generated by loop effects~\cite{ggn} 
but may also be present at the classical level. Thus, we can treat its 
coefficient as a free parameter. However, to be consistent with 4d 
supergravity (which we of course require although, at the technical 
level, the present paper uses only rigid SUSY), the coefficients at the 
two boundaries are assumed to sum up to zero. As discussed in detail 
in~\cite{fit}, the FI terms induce the desired constant bulk VEV of the 
scalar adjoint $\Phi$.

Alternatively, the 5d model may be considered as the small-$R_6$ limit of
a 6d theory, in which case the $\Phi$-VEV corresponds to a Wilson line
wrapping the 6th dimension. It is stabilized by the boundary conditions at
the conical singularities of the 6d model. A more detailed discussion will
be provided in the next section.\footnote{
This 
alternative possibility is interesting in view of the following possible 
criticism of FI-term-stabilization: As argued in~\cite{bccrs}, the FI-terms 
can be understood in supergravity as arising from a mixed gauge-graviphoton
CS term in the bulk. However, in the constructions considered here the brane 
U(1) arises from a non-Abelian gauge symmetry in the bulk and we are not 
aware that the required mixed CS term has been discussed in this case.}

In the above setting, the 4d gauge couplings observed just below the 
compactification scale $M_c=1/R$ read
\beq
\frac{1}{g_{4,i}^2(M_c)}=\frac{\pi R}{g_{5,i}^2}+\frac{1}{g_{bd.,i}^2}\,.
\eeq
Here the 5d gauge couplings are defined at zero momentum (i.e., as in the 
low-energy effective action of Sect.~\ref{prep}) and the last term 
accounts for the (presumably sub-dominant) effect of boundary gauge-kinetic 
terms. From the results of the last two sections, it is now clear that
power-law corrections to inverse 4d gauge couplings are of the order 
$\sim|\Phi|R$ and can thus be as large as the tree-level term $\sim R/
g_{5,cl.}^2$. 

To be more specific, we focus on the situation where $G\!=\,\,\,$SU(5) and 
$P\!=$ diag$(1,1,1,-1,-1)$ so that the breaking is to the SM gauge group. 
In this case, the above $S^1/Z_2$ orbifold of a pure 5d SYM theory gives, 
at the zero mode level, the SM gauge multiplet and a chiral superfield with 
the quantum numbers of the $X,Y$ gauge bosons. The latter one becomes 
massive when $\Phi$ develops a VEV and is therefore phenomenologically 
harmless. SM matter and Higgs fields can be added at the branes and/or in 
the bulk making the model as realistic (and arguably even somewhat simpler 
and more generic) as the more widely discussed $S^1/(Z_2\times Z_2')$ 
models of~\cite{orb5a,orb5b}.

\subsection{Power-law corrections and consistency with boundary\\ anomaly 
cancellation}
For the purpose of this subsection, we treat boundary gauge-kinetic terms 
and the corresponding logarithmic running as sub-dominant. Thus, our 
analysis is based entirely on Eq.~(\ref{gct}), where the gauge group is 
SU(5) and we consider the possibility of hypermultiplet matter in the 
${\bf 5}$, ${\bf 10}$ and ${\bf 24}$. (Recall that, for example, a 
hypermultiplet in the ${\bf 5}$ contains, in 4d ${\cal N}\!=\!1$ language, 
one 4d chiral superfield in the ${\bf 5}$ and one in the $\bar{\bf 5}$.) All 
the group theory we need is the familiar decomposition of the simplest 
SU(5) representations under SU(3)$\times$SU(2)$\times$U(1):
\bea
{\bf 5}  &=& ({\bf 3},{\bf 1})_{-2}+({\bf 1},{\bf 2})_3\\
{\bf 10} &=& ({\bf 3},{\bf 2})_1+(\bar{\bf 3},{\bf 1})_{-4}
+({\bf 1},{\bf 1})_6\\
{\bf 24} &=& ({\bf 8},{\bf 1})_0+({\bf 1},{\bf 3})_0+({\bf 1},{\bf 1})_0+
({\bf 3},{\bf 2})_{-5}+(\bar{\bf 3},{\bf 2})_5\,.
\eea
The U(1) charges $ q'$ given here, in the conventions of~\cite{sla}, 
correspond to charges $q=q'/\sqrt{60}$ if the U(1) generator is normalized 
consistently with the other SU(5) generators. For easy reference we also
collect in Table~\ref{su5} the relevant group-theoretical factors 
$T_{r_i(f)}$ in a hopefully self-explanatory notation.

\begin{table}[ht]
\[
\begin{array}{|ccr|c|c|c|c|}
\hline
 &&& T^{\mbox{\scriptsize U(1)}} & T^{\mbox{\scriptsize SU(2)}} & 
T^{\mbox{\scriptsize SU(3)}} & \mbox{VEV-induced mass}\\
\hline
({\bf 3},{\bf 1})&\quad\mbox{of}\quad&{\bf 5} & 1/5 & 0 & 1/2 & -(2/5)M_V \\
({\bf 1},{\bf 2})&\quad\mbox{of}\quad&{\bf 5} & 3/10 & 1/2 & 0 & (3/5)M_V \\
\hline
({\bf 3},{\bf 2})&\quad\mbox{of}\quad&{\bf 10} & 1/10 & 3/2 & 1 & (1/5)M_V \\
(\bar{\bf 3},{\bf 1})&\quad\mbox{of}\quad&{\bf 10} & 4/5 & 0 & 1/2 & -(4/5)M_V
\\
({\bf 1},{\bf 1})&\quad\mbox{of}\quad&{\bf 10} & 3/5 & 0 & 0 & (6/5)M_V \\
\hline
({\bf 8},{\bf 1})&\quad\mbox{of}\quad&{\bf 24} & 0 & 0 & 3 & 0 \\
({\bf 1},{\bf 3})&\quad\mbox{of}\quad&{\bf 24} & 0 & 2 & 0 & 0 \\
({\bf 1},{\bf 1})&\quad\mbox{of}\quad&{\bf 24} & 0 & 0 & 0 & 0 \\
({\bf 3},{\bf 2})+(\bar{\bf 3},{\bf 2})&\quad\mbox{of}\quad&{\bf 24} & 5 & 3 
& 2 & M_V \\
\hline
\end{array}
\]
\refstepcounter{table}\label{su5}

\vspace*{-.2cm}
{\bf Table~\ref{su5}:} Group-theoretical factors $T_{r_i(f)}$ of the 
simplest SU(5) representations relevant for the evaluation of 
Eq.~(\ref{gct}). The last column contains the masses which the different 
representations acquire in the presence of a $\Phi$-VEV (parameterized by 
the mass $M_V$ of the 5d $X,Y$ gauge bosons).
\end{table}

We are now in a position to write down explicitly the corrections $\Delta 
\alpha_1^{-1}$, $\Delta\alpha_2^{-1}$, and $\Delta\alpha_3^{-1}$ to the 
inverse couplings of the three SM gauge groups U(1), SU(2), and SU(3)
(as usual, $\alpha_1=g_1^2/(4\pi)$ etc.). For example, a ${\bf 5}$ 
hypermultiplet with bulk mass $m_{\bf 5}$ (parameterized by $\xi_{\bf 5}= 
m_{\bf 5}/M_V$) induces corrections 
\bea
&\Delta \alpha_1^{-1}=-\left(\frac{1}{5}\left|\xi_{\bf 5}-\frac{2}{5}\right|+
\frac{3}{10}\left|\xi_{\bf 5}+\frac{3}{5}\right|\right)\,\frac{M_V}{2M_c}
\,\,\,,&\nonumber\\ \label{dac}\\
&\Delta \alpha_2^{-1}=-\frac{1}{2}\left|\xi_{\bf 5}+\frac{3}{5}\right|
\frac{M_V}{2M_c}\qquad,\qquad \Delta \alpha_3^{-1}=-\frac{1}{2}\left|
\xi_{\bf 5}-\frac{2}{5}\right|\frac{M_V}{2M_c}\,\,\,.&\nonumber
\eea
This and corresponding formulae for the ${\bf 10}$ hypermultiplet and the 
${\bf 24}$ hypermultiplet or vector multiplet are easily read off from 
Table~\ref{su5} and Eq.~(\ref{gct}) after compactification on an interval 
with length $\pi R=\pi/M_c$. 

Finally, we need to deal with the effect of a classical SUSY CS
term parameterized by $c_{cl.}$. This term is constrained by boundary 
anomaly cancellation~\cite{agw,ch}. As can be seen explicitly from 
Eqs.~(\ref{lc}) and (\ref{pro}), a bulk ${\bf 5}$ in the limit $m_{\bf 5}
\to\pm\infty$ induces an effective CS term with $c_{cl.}=\mp 1/2$~\cite{ims}. 
The boundary anomalies induced by this term can be found as follows (see, 
e.g.,~\cite{bccrs,sssz}):

Consider first two massless bulk hypermultiplets ${\bf 5}$ and ${\bf 5}'$, 
each with the same boundary conditions at $x^5=0$ and $x^5=\pi R$, but with 
the sign flipped between the two hypermultiplets. The model is anomaly-free, 
not just at the zero-mode level but also at each of the two boundaries 
taken separately. This is clear since the zero-mode matter is vector-like, 
so that there is no 4d anomaly, and the boundary anomaly is simply 1/2 of
the 4d anomaly. (Recall that there are no anomalies in 5d.) Furthermore, 
the consistency is not destroyed by continuously varying one of the mass 
parameters, e.g., taking $m_{\bf 5}\to\infty$ while keeping $m_{{\bf 5}'}=0$. 

Thus, the CS term induced by the infinitely heavy ${\bf 5}$ 
precisely cancels the boundary anomalies coming from brane localized 
zero-modes emerging in the limiting procedure $m_{\bf 5}\to\infty$ and 
from the massless ${\bf 5}'$. The latter are half-integer-valued in units 
corresponding to a 4d chiral fermion in the ${\bf 5}$. This is obvious 
since, again, the zero-mode anomaly is split equally between the two 
identical boundaries. Postponing a more explicit discussion to the next 
subsection, we can now already conclude that $c_{cl.}=\mp 1/2$ induces 
half-integer boundary anomalies. Thus, in the absence of charged bulk 
matter, $c_{cl.}$ must be integer and, to achieve gauge invariance, 
appropriate brane fields cancelling the induced integer-valued anomalies 
must be present. This argument for the value of $c_{cl.}$ could have also
been made on the basis of the $m_{\bf 10}\to\infty$ limit of a ${\bf 10}$ 
hypermultiplet, which induces a CS term identical to that induced by a 
${\bf 5}$. 

From Eq.~(\ref{dac}) and corresponding formulae for the matter in the 
${\bf 10}$ and ${\bf 24}$, it is clear that almost any ratio of low-energy 
gauge couplings can be realized by tuning appropriately the bulk masses of 
the matter fields. We therefore now focus on the arguably more natural 
case where bulk fields are either massless or extremely heavy, i.e., 
contribute only via an analytic CS term. The relevant contributions to 
the differences of inverse 4d gauge couplings $\alpha_{ij}= 
\alpha_i^{-1}-\alpha_j^{-1}$ are given in Table~\ref{da}. 

\begin{table}[ht]
\[
\begin{array}{|l|c|c|}
\hline
\mbox{massless fields or operator}& \alpha_{12}\times 2M_c/M_V & 
\alpha_{23}\times 2M_c/M_V \\
\hline
{\bf 24} \,\,\mbox{vector} & 2 & 1 \\
\hline
{\bf 24} \,\,\mbox{hypermultiplet} & -2 & -1 \\
\hline
\,\,\,{\bf 5} \,\,\mbox{hypermultiplet} & 1/25 & -1/10 \\
\hline
{\bf 10} \,\,\mbox{hypermultiplet} & -27/25 & 3/10 \\
\hline
\mbox{CS term with}\,\,\, c_{cl.}=\mp 1/2 & \pm 1/5 & \mp 1/2 \\
\hline
\end{array}
\]
\refstepcounter{table}\label{da}

\vspace*{-.2cm}
{\bf Table~\ref{da}:} Corrections to inverse gauge coupling differences 
(in units of $M_V/(2M_c)$) induced by massless fields in the simplest 
representations and by the smallest possible CS terms.
\end{table}

At this point, some basic phenomenological implications can already be 
derived. Note first that anomaly cancellation by boundary fields is only 
possible if the boundary anomalies induced by bulk fields and operators 
are integer-valued. Thus, the sum of the numbers of bulk ${\bf 5}$s, 
${\bf 10}$s and ``CS-term-quanta'' (i.e., CS term contributions with 
$c_{cl.}=\pm 1/2$) has to be even. 

Recall that 4d MSSM running gives $\alpha_{12}/\alpha_{23}=
7/5=1.4$, which is known to agree very well with the 
observed low-energy gauge couplings. The effect of just the gauge sector
gives, both in the 4d logarithmic and in the above power-law case, 
$\alpha_{12}/\alpha_{23}=2$. In the 4d case, this is then corrected 
by the contribution from the two Higgs doublets. As noted in~\cite{hw}, 
a single bulk hypermultiplet in the ${\bf 5}$, with $m_{\bf 5}$ tuned such 
that, in the presence of the $\Phi$-VEV, the doublet is massless in 5d, 
reproduces the approximately correct ratio $\alpha_{12}/\alpha_{23}=1.2$ 
of~\cite{ddg}. However, as the anomaly argument above 
shows, such a single bulk ${\bf 5}$ has to be supplemented with a CS term.
Unfortunately, this destroys the approximately correct power-law effect
of~\cite{ddg} (this important point was missed in~\cite{hw}). 

Now, coming back to the more restrictive framework of Table~\ref{da}, we
can look for simple configurations which give the MSSM prediction of 
$\alpha_{12}/\alpha_{23}=7/5$ as a power-law effect. It is 
interesting to observe that, indeed, the vector multiplet together with a 
massless bulk ${\bf 10}$ and the minimal required CS term (choosing the 
negative sign, $c_{cl.}=-1/2$) 
gives precisely $\alpha_{12}/\alpha_{23}=7/5$. Thus, this 
combination of bulk fields and operators generates a power-law effect
mimicking MSSM 1-loop running. Furthermore, replacing the ${\bf 10}$ 
with a ${\bf 5}$ and changing the sign of the CS term, one finds 
$\alpha_{12}/\alpha_{23}=46/35\simeq 1.31$, which is also quite close 
to the desired value $1.4$. For the moment, we are satisfied with the
two above examples and leave it to the reader to explore other, more 
complicated, matter field and CS term configurations. We believe that such
a complete analysis should be performed in a more constrained context, e.g.,
in the search for a realistic flavour model or in the framework of a 
first-principles string construction.

\subsection{Low-energy field content}
It remains to be shown how, on the basis of a given bulk matter content and 
CS term, a full anomaly free model is built. We illustrate this construction 
using the particularly attractive scenario with SU(5) vector multiplet, 
${\bf 10}$ hypermultiplet and CS term with $c_{cl}=-1/2$ in the bulk, where 
the power law effect is equivalent to logarithmic MSSM running.\footnote{
It 
is interesting to speculate that this field content arises from a
(possibly even-higher-dimensional) SO(10) model where the adjoint 
decomposes as ${\bf 45}={\bf 24}+{\bf 10}+\overline{\bf 10}+{\bf 1}$ and the 
$\overline{\bf 10}$ becomes heavy in the process of gauge-symmetry and SUSY 
breaking.
}

As before, we compactify on $S^1/Z_2$ breaking SU(5) to the SM at both 
boundaries. We start with no bulk CS term but with a ${\bf 10}$ and 
${\bf 10}'$ bulk hypermultiplet with opposite boundary conditions. In this 
situation, the spectrum of fermionic fields which are non-zero at any of 
the two boundaries is vector-like, i.e., no boundary anomalies arise. By 
continuity, the consistency of this model is not destroyed if, while keeping 
$m_{\bf 10}=0$, the limit $m_{\bf 10'}\to\infty$ is taken. We now have an 
anomaly-free model with the desired content of light bulk fields and a CS 
term with $c_{cl}=-1/2$. A specific brane field content arises from the 
${\bf 10}'$ in the limit $m_{\bf 10'}\to\infty$ due to the presence of 
localized zero-modes~\cite{jr} (see also~\cite{ad}). 

To discuss these brane fields explicitly recall that, in ${\cal N}\!=\!1$ 
language, the ${\bf 10}'$ hypermultiplet contains two chiral superfields in 
complex-conjugate representations, which we denote by ${\bf 10}'$ and 
${\bf 10}'^c$. Assume that the sign-conventions of the 5d lagrangian are 
such that positive $m_{\bf 10'}$ implies a localization\footnote{
By
this we mean that the relevant bulk equations of motion for the ${\cal N}\!
=\!1$ superfields are $(\partial_y+m_{\bf 10'}){\bf 10'}=0$ and $(\partial_y
-m_{\bf 10'}){\bf 10'}^c=0$, implying bulk solutions ${\bf 10'}\sim\exp(-
m_{\bf 10'}y)$ and ${\bf 10'}^c\sim\exp(+m_{\bf 10'}y)$. In particular, if 
a zero mode is allowed by the boundary conditions, it will then be 
localized as described in the main text above. 
}
of the ${\bf 10}'$ 
at $y=0$ and of the ${\bf 10}'^c$ at $y=\pi$. Furthermore, we define the 
SU(5)-breaking boundary conditions such that the $({\bf 3,2})'$ is 
non-zero while the $({\bf\bar{3},1})'$ and $({\bf 1,1})'$ vanish at both 
branes. It is now clear that, in the limit $m_{\bf 10'} \to\infty$, the only 
light fields are the zero mode of $({\bf 3,2})'$, completely localized 
at $y=0$, and the zero modes $({\bf\bar{3},1})'^c$ and $({\bf 1,1})'^c$, 
completely localized at $y=\pi$. 

Phenomenologically, it also essential to know what zero modes arise from the 
${\bf 10}$ hypermultiplet and at which brane they are peaked. (Note that, in 
contrast to the complete localization of the zero modes arising from the 
${\bf 10}'$ hypermultiplet, we have strong but finite peaking characterized 
by exp$[\pm ym]$.) Given our conventions for the relative sign between bulk 
mass and $\Phi$-VEV, as specified by Eq.~(\ref{dac}), and the signs in the 
last column of Table~\ref{su5}, the direction of the peaking of the various 
fields of the ${\bf 10}$ hypermultiplet is easily determined. Recalling 
that the boundary conditions of the ${\bf 10}$ hypermultiplet are opposite 
to those of the ${\bf 10}'$ hypermultiplet, we find a $({\bf 1,1})$ zero 
mode peaked at $y=0$ as well as $({\bf 3,2})^c$ and $({\bf\bar{3},1})$ 
zero modes localized at $y=\pi$. 

To make the model realistic without destroying the MSSM-like power law
contribution from the bulk, matter has to be introduced in the form of brane 
fields. We begin by localizing a ${\bf 10}$ chiral superfield at $y=\pi$. 
Allowing all gauge-invariant mass terms and recalling that, from the previous 
construction, we also have a $({\bf 3,2})^c$, $({\bf\bar{3},1})$, 
$({\bf\bar{3},1})'^c$ and $({\bf 1,1})'^c$ peaked or localized at 
$y=\pi$, we find that all fields except for a partnerless $({\bf\bar{3}, 
1})_{-4}$  become massive. Together with the $({\bf 3,2})_1$ 
and $({\bf 1,1})_6$ left over at $y=0$ from the previous construction, we
now have a full ${\bf 10}$ in zero modes. Of course, the introduction of a 
${\bf 10}$ chiral superfield at $y=\pi$ demands, by anomaly cancellation, 
the further introduction of a $\bar{\bf 5}$ at the same brane. Now we have 
a full SM generation, with all fields except for the left-handed quarks and 
right-handed electron peaked at $y=\pi$. Amusingly, this matter distribution 
excludes all (not exponentially suppressed) mass terms within this 
generation. Thus, our construction has produced an anomaly-free setting 
with MSSM-like power law correction and one naturally light generation. 
The two Higgs doublets and the heavy generations are now easily added 
at $y=\pi$ without affecting any of the attractive features achieved so 
far. 

To be completely explicit, we now calculate the gauge couplings at the 
$Z$-pole in the above model, including logarithmic terms. Below the 
compactification scale $M_c=1/R$ we have conventional MSSM running; above
that scale we have the power-like effects discussed in this paper and further 
corrections associated with the logarithmic running of 
brane-localized gauge-kinetic terms (see, e.g.,~\cite{nsw,orb5b,cprt,hw}). 
This logarithmic running above $M_c$ is cut off at some UV-scale $\Lambda$ 
where the singular boundary is resolved. For simplicity, we assume 
$\Lambda/M_V={\cal O}(1)$ and thus disregard logarithms of this ratio. In 
particular, this implies that only the Kaluza-Klein (KK) modes of the 5d 
vector multiplet within the SM gauge group contribute to the logarithmic 
running above $M_c$. 

Note that, from the point of view of the bulk theory and the power-like 
terms, the existence of a UV scale $\Lambda$ is immaterial since our 
calculation of inverse gauge coupling differences is entirely 
UV-insensitive. In fact, this was to be expected in 
view of the possible existence of a non-trivial UV fixed-point of the 5d 
theory discussed in~\cite{ims}, i.e., the possibility of taking $\Lambda 
\to\infty$ (see~\cite{fix} for more general analyses, including in 
particular the 6d case, and~\cite{ddg1} for a 
recent application in unified models). However, we emphasize that our 
calculations, although quite consistent with the fixed point proposal, do 
not rely on it or on the limit $\Lambda\to\infty$ since all dangerous 
higher-dimension operators are forbidden by symmetries.

The low-energy inverse gauge couplings are given by
\beq
\alpha_{4,i}^{-1}(m_Z)=\pi R\alpha_{5,cl.}^{-1}+b_i\left(\frac{1}{10}
\frac{M_V}{M_c}+\frac{1}{2\pi}\ln\frac{M_c}{m_Z}\right)+\frac{\tilde{b}_i}
{2\pi}\ln\frac{M_V}{M_c}+\{\mbox{$i$-indep.~terms}\}\label{lr}
\eeq
where $\alpha_{5,cl.}=g_{5,cl.}^2/(4\pi)$. The coefficients 
$b_i=(0,-6,-9)+2\,(3/10,1/2,0)$ govern the familiar gauge
and Higgs contributions to the MSSM running and, in our specific example,
also the power-law term. Their $S^1/Z_2$ counterparts governing the 
modified running above $M_c$ are $\tilde{b}_i=(0,-4,-6)+2\,(3/10,1/2,0)$.
Note that, to simplify Eq.~(\ref{lr}), we have chosen the ``$i$-independent 
terms'' to ensure that the familiar coefficients $b_i$ multiply both the 
power-law term and $\ln(M_c/m_Z)$. This is possible because the
power-law corrections respect the MSSM relation $\alpha_{12}/\alpha_{23}=
7/5$. 

The main technical statement to be made is the harmlessness of this 
modified logarithmic contribution, which is sufficiently similar to MSSM
running and parametrically much smaller than the power law term. To see
this explicitly, consider the most extreme case of $M_c\sim m_Z$ (i.e., 
disregard the term $\sim\ln(M_c/m_Z)$) and choose $M_V=48.5\,M_c$. 
Equation~(\ref{lr}) then gives $\alpha_{12}(m_Z)=29.4$ and $\alpha_{23}
(m_Z)=21.3$ in almost perfect agreement with what is needed to 
accommodate the low-energy values $\alpha_i^{-1}(m_Z)=(59.0,29.6,8.4)$.

\section{Power corrections from 6 dimensions}\label{6d}
In this section, we discuss power-like corrections to gauge unification 
in 6d SYM theories. To begin, consider uncompactified, flat, 6-dimensional 
space with minimal SUSY (corresponding to ${\cal N}\!=\!2$ in 4d), in which 
case the vector multiplet contains just the gauge field and a 6d-chiral 
spinor~\cite{fay}. We may add 6d gauged hypermultiplets, the spinors of 
which must be of opposite 6d chirality relative to the gaugino. The reason
for this is the presence of a Yukawa-like interaction term in the 6d 
lagrangian. This term combines the gaugino with the charged matter fermion, 
forcing them to have opposite chirality. (In this context, it is useful to
recall that, unlike in 4d, in 6d complex conjugation does not change the 
chirality of a spinor. Thus, 6d chirality is an `absolute concept' in the 
sense that it does not depend on whether one views the spinor or its complex 
conjugate as the basic degree of freedom.)

The above implies that no mass terms connecting gauged hypermultiplets are
allowed in 6d. Indeed, all the fermions involved have the same chirality
making fermionic mass terms impossible. Independent of the gauging, the 
absence of masses in 6d simply follows from the fact that, in a 
hypermultiplet, the SUSY variation of a fermion is proportional to the SUSY 
generator, implying that all the fermions have the same chirality. This is 
very interesting from the model building perspective since it implies that 
the 5d hypermultiplet masses, which could in principle be used for an 
arbitrary tuning of 5d power-like unification corrections 
(cf.~Eqs.~(\ref{gc}) and (\ref{dac})) have no 6d analogue. 

However, it would be premature to conclude that there is no massive gauged 
matter in 6d. Indeed, mass terms linking a 6d hypermultiplet with a 6d vector 
multiplet, both charged under some gauge group, are possible. Such mass 
terms arise, for example, in the KK mode description of $d$-dimensional 
theories, where $d\!>\!6$, compactified to 6d. They also appear in 
situations where a 6d gauge symmetry is broken by the VEV of the scalar 
component of one of the gauged hypermultiplets. Mass terms of this type, 
involving vector and hypermultiplet in the same representation, 
automatically produce a full ${\cal N}\!=\!4$ multiplet at a given mass 
level. Thus, they are irrelevant in the present context of loop corrections 
to gauge coupling unification and we can from now on focus on massless 6d 
models.\footnote{
More 
generally, according to Tables 4 and 5 of~\cite{stra} all massive 
representations of minimal 6d SUSY with spin $\le\!1$ automatically have 
a 4d ${\cal N}\!=\!4$ spectrum.}

The 6d vector multiplet contains no scalar (the adjoint $\Phi$ of the 
corresponding 5d theory being promoted to the gauge field component 
$A_6$). Thus, soft gauge symmetry breaking in a 6d Lorentz-invariant 
setting has to rely on the VEV of one of the scalars of a gauged 
hypermultiplet. As explained above, massive fields can be collected 
in full ${\cal N}\!=\!4$ SUSY multiplets for any given mass and 
representation and no power corrections to gauge unification arise. 
This ends our discussion of the uncompactified 6d theory. What is more, it 
also implies that the only interesting situation in 5d is the one where the 
gauge symmetry breaking is driven by the adjoint scalar from the vector
multiplet. Indeed, a 5d theory broken by a hypermultiplet VEV can be 
thought of as arising via dimensional reduction from a 6d theory, in which
case the above argument demonstrates the absence of power-like loop 
corrections. This is the reason why our 5d analysis is focussed entirely 
on situations with gauge symmetry breaking by the adjoint scalar $\Phi$. 
It may, however, be interesting to consider situations where bulk 
hypermultiplet VEVs are present in addition to the VEV of the adjoint 
scalar. 

Given the absence of power-law corrections in the Lorentz-invariant 6d
situation, we now focus on 6d theories compactified on an $S^1$ of radius 
$R_6$ to 5 dimensions. Any possible further compactification (with 
compactification radius $R_5$) leading to a realistic 4d model is assumed 
to occur at a lower energy scale, $R_5\gg R_6$. In the 5d effective 
theory, the gauge symmetry can be broken by the VEV of the adjoint scalar 
$\Phi$. The latter has to be identified with the VEV of $A_6$, i.e., the 
Wilson line wrapping the $S^1$~\cite{hos}. Thus, one can straightforwardly apply the 
analysis of the previous sections and obtain the power-law corrections for 
any given 6d model. Important new features are the absence of a classical CS 
term and of hypermultiplet masses in 6d, which makes the setting more 
predictive, and the appearance of a tower of KK modes, the loop contributions 
of which have to be summed. The remainder of this section is devoted to a 
detailed discussion of power-law effects in this effectively 5-dimensional 
situation. 

Before coming to the actual calculation, another conceptual issue -- the 
stabilization of the Wilson line -- has to be addressed. For the simplest 
geometrical setting, a rectangular torus $T^2$ with radii $R_5$ and $R_6$, 
the Wilson line in $x^6$-direction, which is the analogue of the $\Phi$-VEV
of the 5d models above, is a modulus protected by SUSY. However, in an
appropriate orbifold of the type $T^2/Z_2$, $T^2/(Z_2\times Z_2')$ etc.,
the Wilson lines have certain fixed, discrete values determined by the gauge
twists associated with the various orbifold actions~\cite{orb6}. In fact, 
this is quite analogous to the discrete or quantized Wilson lines of 
string-theoretic orbifold models~\cite{dhvw}. To be specific, recall that a 
$T^2/Z_2$ orbifold can be visualized as the surface of a 
`pillow'~\cite{abc1}. It has the topology of a sphere and 4 conical 
singularities with deficit angle $\pi$. In various field- or string-theoretic 
orbifold constructions, gauge symmetry breaking on this space arises from 
the non-trivial gauge holonomy associated with loops surrounding the 
`corners' of this pillow. By Gauss' theorem, two of these Wilson lines 
surrounding two adjacent corners combine into a Wilson line going around the 
center of the pillow, which will therefore in many cases have a non-zero, 
quantized value. It is now straightforward to imagine an extremely elongated 
pillow ($R_5\gg R_6$) equipped with a fixed Wilson line in $x^6$ direction. 
The conical singularities are simply boundary effects (from the effective 5d
point of view) stabilizing the Wilson line VEV. In fact, as discussed 
in~\cite{hr}, in field theory the Wilson lines surrounding each of the 
conical singularities do not have to be quantized but can vary continuously
and each possible value can be stabilized by local physics at the fixed 
point (brane). An example for such a local stabilization mechanism is 
provided by brane-localized FI terms inducing locally a non-zero field 
strength~\cite{lnz}.

\subsection{Power corrections in a 6d theory compactified on a circle}
\label{cir}
We have now identified an interesting and realistic setting for 6d power
corrections: the effectively 5-dimensional case with a Wilson line wrapping 
the compactified dimension of length $2\pi R_6$. In fact, the corrections to 
gauge coupling unification arising in this setup could be extracted from the 
more general analysis of arbitrary tori with two Wilson lines performed 
in~\cite{gng,ghi} (including a discussion of the connection to string 
theory~\cite{bac}). Terms linear in the Wilson line VEV appear, for example, 
in taking 
the appropriate limits of Eq.~(27) in~\cite{ghi}. However, we find it useful 
to give an independent and extremely simple derivation, based on the 5d 
results obtained 
above, which adequately describes the dominant part of large, power-like 
corrections to gauge unification. It is important to note that the 
analysis of~\cite{gng} supports the expectation (based, e.g., on the 
symmetry arguments or the UV fixed-point conjecture of~\cite{ims}) that 
the field theory results for gauge coupling differences are recovered in 
string theory in the limit of infinite string tension. 

Consider first, as at the beginning of Sect.~\ref{pl}, a supersymmetric 6d 
U(1) gauge theory with a gauged hypermultiplet of charge $q$. This will 
be a useful building block for the following realistic calculation although, 
without appealing to the Green-Schwarz mechanism, the simple U(1) model is 
inconsistent since it is anomalous.\footnote{
For 
a 6d U(1) model, the anomaly induced by the box diagram is always non-zero 
since it is proportional to the sum of the fourth powers of the charges of 
the fermions, which all have the same chirality because of 6d SUSY. This 
will be different in non-abelian models (see below) since the gaugino is
charged and has opposite chirality.
}
After compactification, we have a KK tower of 5d hypermultiplets with masses 
$m_n=|n/R_6|$ with $n$ integer. Turning on a Wilson line in $x^6$ direction, 
the former zero mode acquires a non-zero mass $m=qA_6$ (where we have 
chosen a gauge with constant $A_6$-VEV). A corresponding Wilson-line-induced 
effective mass correction is also added to the masses of the higher KK modes 
(which is particularly evident in the fermionic part of the lagrangian). 
The resulting KK spectrum is $m_n=|n/R_6+m|$ with $n$ running over all 
integers. Thus, the loop correction of Eq.~(\ref{lo}) is replaced by
\beq
\delta\left(\frac{1}{g_5^2}\right)=-\frac{q^2}{8\pi^2}
\sum_{n=-\infty}^{+\infty}\,|nR_6^{-1}+m|\,.\label{wls}
\eeq
As before, we are only interested in the mass dependence of this correction.
This mass dependence is finite and can be easily extracted from the above 
divergent sum using dimensional regularization. It is convenient to introduce 
the dimensionless parameter $c=mR_6=qA_6R_6$ assuming $0\!<\!c\!<\!1$ for 
the moment. The result, derived in Appendix~B, then reads
\beq
\delta\left(\frac{1}{g_5^2}\right)=-\frac{q^2}{8\pi^2R_6}\,c(1-c)
=-\frac{q^2}{8\pi^2}\,m\,(1-c)\,,\label{6du1}
\eeq
where, we emphasize again, an $m$-independent divergent contribution has
been dropped. 

\begin{figure}[ht]
\begin{center}
\includegraphics[width=7cm]{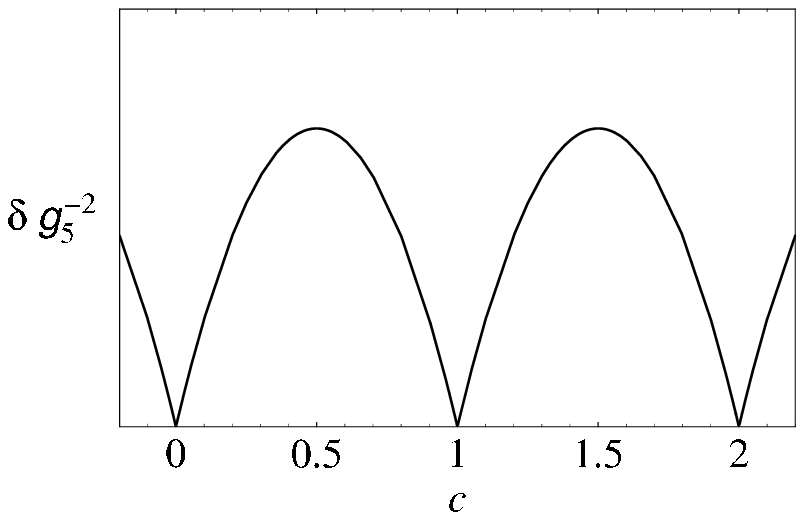}
\end{center}
\refstepcounter{figure}\label{u1}

\vspace*{-.2cm}
{\bf Figure~\ref{u1}:} Illustration of the dependence of the 1-loop 
correction to the inverse gauge coupling in the 5d effective theory, 
$g_5^{-2}$, on the value of the Wilson line, parameterized by $c=qA_6R_6$. 
\end{figure}

This very simple formula has manifestly the correct limiting behaviour as 
$R_6\to 0$ for fixed $m$. Furthermore, viewed as a function of $R_6$ and 
$c$, it is invariant under the substitution $c\to(1\!-\!c)$. This is a 
manifestation of the fact that the KK spectrum is completely determined 
once the lightest mode is known. In other words, the point $n=0$ has no 
absolute meaning and a shift of the label $n$ or a reflection $n\to -n$ 
do not affect the physics. This last comment makes it obvious that 
Eq.~(\ref{6du1}) is extended to values of $c$ outside the interval $(0,1)$ 
by simply demanding reflection symmetry with respect to any point where $c$ 
is integer. This is illustrated in Fig.~\ref{u1}. 

It is evident from Fig.~\ref{u1} that, locally, the inverse gauge coupling
squared depends quadratically on the Wilson line VEV and thus, from the 
5d point of view, on $\Phi$. However, we know from SUSY and gauge invariance 
(cf. Sect.~\ref{prep} and Appendix~A) that the 5d prepotential is
at most cubic and thus the $\Phi$ dependence is at most linear. This 
inconsistency is directly linked to the non-zero anomaly, as can be easily 
seen from Eq.~(\ref{6du1}). Indeed, for a model with several hypermultiplets 
the term quadratic in $A_6$ is proportional to the sum of the fourth
powers of the charges, i.e., the anomaly, which is necessarily non-zero. 
We will see shortly that this problem disappears in an anomaly free, 
non-abelian model.

The non-abelian version of the above result can be written down without any 
further calculation. Recall that, at the beginning of Sect.~\ref{pl}, we 
have given a rederivation of Eq.~(\ref{gc}) on the basis of Eq.~(\ref{lo}) 
and simple group theory. Following this line of reasoning, the 6d version of 
Eq.~(\ref{gc}) can now immediately be given:
\bea
{\cal L}_i &\supset &-\frac{1}{4}F_i^2\left\{\frac{2\pi R_6}{g_{6,cl.}^2}+
\frac{1}{8\pi^2}\left(\sum_\alpha\,\alpha_i^2\,|\alpha_jA_6^j
|\,\Big(1-|\alpha_jA_6^j|R_6\Big)\right.\right.\label{nali}
\\
&&\hspace*{4cm}\left.\left. -\sum_f\sum_\lambda\,\lambda_i^2\,|\lambda_j
A_6^j|\,\Big(1-|\lambda_jA_6^j|R_6\Big)\right)\right\}\,.\nonumber
\eea
It is obtained from the original expression by identifying each 5d mass $m$ 
and replacing it by $m(1-mR_6)$. This is the same procedure that leads 
from Eq.~(\ref{lo}) to its 6d version Eq.~(\ref{6du1}). Of course in 
addition, the components $\phi^i$ of the field $\Phi$ are replaced by the 
corresponding components $A_6^i$ of $A_6$ and the classical CS term as well 
as the hypermultiplet masses are dropped. 

Similarly, the 6d analogue of Eq.~(\ref{gct}), which is most directly 
useful for GUT phenomenology, reads
\beq
\delta\left(\frac{1}{g_{5,i}^2}\right)=\frac{1}{8\pi^2}\left(\sum_{r_i(a)}
T_{r_i(a)}M_{r_i(a)}\Big(1\!-\!M_{r_i(a)}R_6\Big)-\sum_f\sum_{r_i(f)}
T_{r_i(f)}M_{r_i(f)}\Big(1\!-\!M_{r_i(f)}R_6\Big)\right)\,.\label{nade}
\eeq

It is clear from the structure of Eq.~(\ref{nali}) that each of the 5d 
low-energy $U(1)$ gauge couplings depends on $A_6$ like a sum of functions 
of the type displayed in Fig.~\ref{u1}. In fact, both Eq.~(\ref{nali}) and 
Eq.~(\ref{nade}) can be taken at face value only in a certain neighbourhood 
of the point $A_6=0$. They are extended to all values of $A_6$ along a 
certain direction in the Cartan subalgebra by extending each of the terms
of the form $m(1-mR_6)$ as illustrated in Fig.~\ref{u1}. Locally, the sum 
of these terms must be a linear function since the 5d prepotential is at 
most cubic. The required cancellation of the coefficient of $(A_6)^2$ is 
indeed possible because of the relative sign between the vector multiplet 
and the hypermultiplet contributions in Eq.~(\ref{nali}). This 
cancellation is intimately linked to the absence of 6d anomalies. To see
this more explicitly, let $A_6^i$ (with $i$ fixed) be the only non-zero 
component of $A_6$ and consider the gauge coupling correction to the 
U(1) subgroup generated by $H_i$ as specified by Eq.~(\ref{nali}). The 
coefficient of $(A_6^i)^2$ is now manifestly proportional to the box
anomaly coefficient. It vanishes whenever the sum of the fourth powers of 
charges (specified by $\alpha_i$) of fermions of the gaugino-chirality minus 
the sum of the fourth powers of charges (specified by $\lambda_i$) of
fermions of matter-chirality is equal to zero. This is a nice consistency 
check of the present analysis.

\subsection{A 6d SO(10) example}
As an illustration of the general discussion above we now explicitly 
calculate the power corrections to gauge unification in a 6d SO(10) model 
compactified to 5d on an $S^1$. The group is broken to SU(3)$\times$SU(2)$
\times$U(1)$\times$U(1)$'$ by an $A_6$ Wilson line along the hypercharge 
direction, which corresponds to the first of the two U(1)s above. (For a 
detailed discussion of the corresponding group theory and the various 
breaking possibilities see, e.g., Sect.~3.2 of~\cite{hr}.) 

One possible special case of 5d effective theories of this type arises in 
the orbifold models of~\cite{orb6}. These models have a pillow-like 
fundamental space with gauge symmetries SO(10), SU(5)$\times$U(1), 
SU(5)$'\times$U(1)$'$ and SU(4)$\times$SU(2)$\times$SU(2)) at the four 
corners. One can now imagine stretching this space in one direction such 
that the SO(10) and the Pati-Salam fixed points are at one of the two 
boundaries of the resulting effectively 5-dimensional model (while the SU(5) 
and the flipped SU(5) fixed points are at the other boundary). Away from the 
boundaries, we have a cylinder wrapped by a Wilson line in 
hypercharge-direction, which is precisely the 6 to 5d compactification 
discussed above. In this specific orbifold realization, the Wilson line is 
quantized such that it corresponds to a $Z_2$ gauge twist and 
correspondingly the gauge symmetry in the 5d bulk is enhanced from the generic
case, SM$\times$U(1)$'$, to the Pati-Salam group (cf. the 5d 
the models of~\cite{pst}). However, one can clearly imagine other similar 
constructions with different values of the $A_6$ Wilson line (see, e.g., 
the models of~\cite{hr} and~\cite{lnz} where Wilson lines encircling 
conical singularities take on continuous values not related to the 
geometrical deficit angle).

Restricting ourselves to hypermultiplet matter in the ${\bf 10}$ and 
${\bf 16}$ of SO(10), there is only one model without irreducible or 
reducible gauge anomalies. It contains, in addition to the vector multiplet 
in the ${\bf 45}$, 6 hypermultiplets in the ${\bf 10}$ and 4 hypermultiplets 
in the ${\bf 16}$ of SO(10)~\cite{abcf}. The existence and uniqueness of 
this solution is easily checked using the formulae of~\cite{hmrs} (based 
on~\cite{aw} and~\cite{schw}). More possibilities exist if one only 
requires that the irreducible anomaly cancels, appealing to the 
Green-Schwarz mechanism~\cite{gs} for the cancellation of the reducible 
anomalies. We leave the investigation of power-law corrections in this 
context to future work. We also do not discuss 4d boundary anomalies arising 
at the conical singularities of the full model~\cite{abca} since they are
not an intrinsic part of the effective 5d theory in which the 
power-corrections arise. However, we emphasize that an example of a 
realistic SUSY GUT with the above anomaly-free 6d bulk matter content 
has been given in~\cite{abcf}. 

In principle, the calculation of the power-law corrections in the 
anomaly-free 6d SO(10) model is a straightforward application of 
Eq.~(\ref{nade}). The analysis becomes particularly simple if one uses 
the 5d results of Table~\ref{su5} together with the familiar decomposition
of SO(10) representations in SU(5) language. Specifically, the matter 
content of a vector ${\bf 45}$ and hypermultiplets $6\times{\bf 10}+4\times 
{\bf 16}$ of SO(10) corresponds to vector multiplets ${\bf 24}+
2\times{\bf 10}$ and hypermultiplets $16\times{\bf 5}+4\times{\bf 10}$ of 
SU(5). (Note that, as far as gauge coupling corrections are concerned, we 
do not need to distinguish between ${\bf 5}$ and $\bar{\bf 5}$ etc.)
The effective 5d masses of the various fields are completely specified by 
these SU(5) representations since the symmetry-breaking Wilson line lies 
within the SU(5) subgroup. In particular, the two vector ${\bf 10}$s cancel 
the effect of two of the hypermultiplet ${\bf 10}$s of SU(5) because of
effective ${\cal N}\!=\!4$ SUSY in the spectrum. 

\begin{figure}[ht]
\begin{center}
\includegraphics[width=12cm]{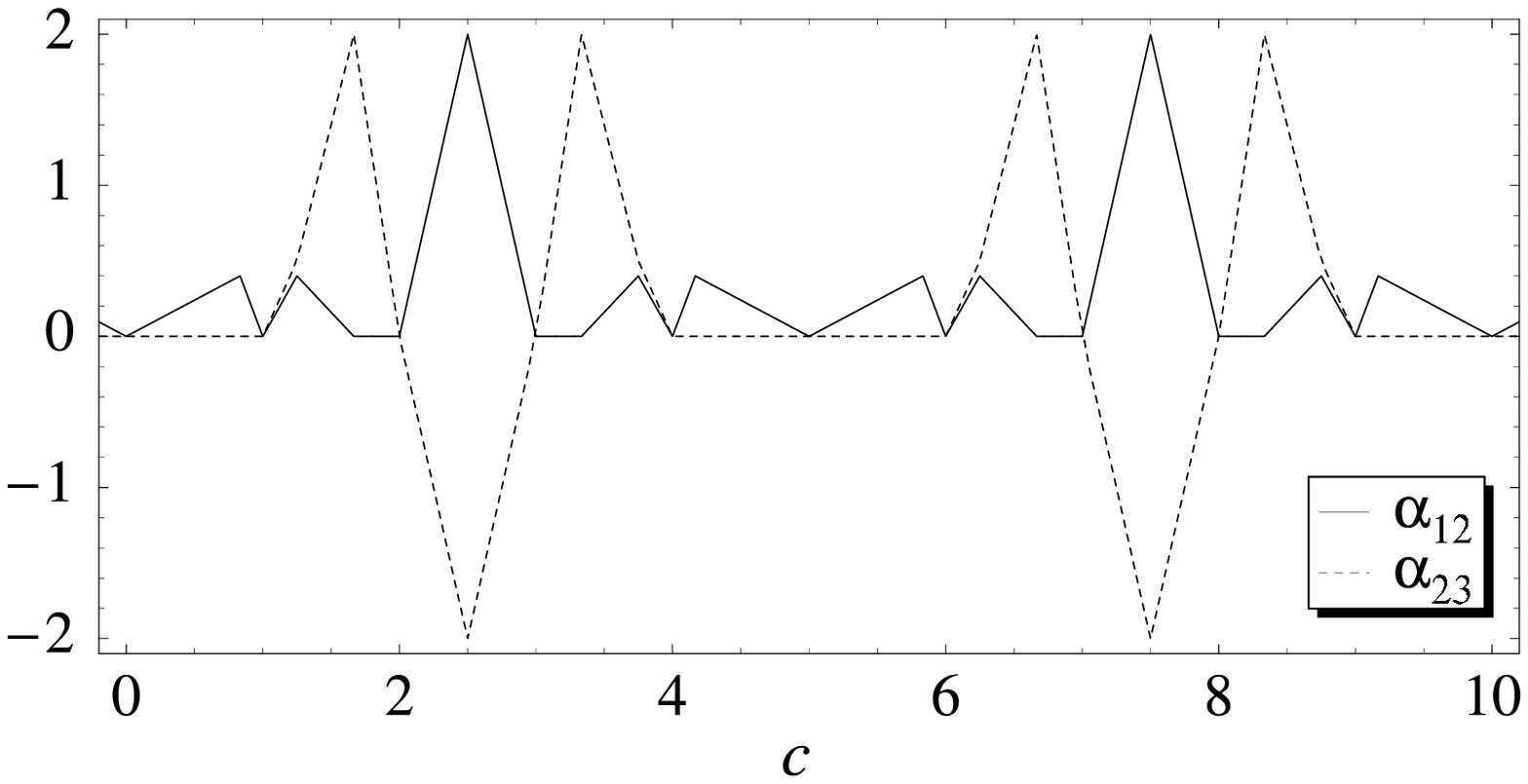}
\end{center}
\refstepcounter{figure}\label{so10}

\vspace*{-.2cm}
{\bf Figure~\ref{so10}:} Power corrections to the inverse gauge coupling
differences $\alpha_{12}$ and $\alpha_{23}$ in units of $(R_5/2R_6)$ as
functions of $A_6$ (parameterized by $c=M_VR_6$, where $M_V$ is the 6d 
$X,Y$ gauge boson mass). Note the piecewise linear form related to 6d anomaly
cancellation.
\end{figure}

Of course, the modification of corrections of the type displayed in 
Eq.~(\ref{dac}) arising from the summation of the full KK tower has to be 
taken into account as described in Sect.~\ref{cir}. For example, the fields 
of one 6d bulk hypermultiplet in the ${\bf 5}$ give a correction
\beq
\Delta\alpha_2^{-1}=-\frac{1}{2}\left(\frac{3}{5}M_V\right)\left(1-\frac{3}
{5}M_VR_6\right)\frac{R_5}{2}+\cdots=-\frac{1}{2}\left(\frac{3}{5}c\right)
\left(1-\frac{3}{5}c\right)\frac{R_5}{2R_6}+\cdots\,,
\eeq
where $c=M_VR_6$. Here we have assumed that the effective 5d theory is further
compactified to 4d on an interval of length $\pi R_5$ following as closely 
as possible the purely 5-dimensional situation of Sect.~\ref{5d}. As before, 
the typical $A_6$ dependence arising from the structure $m(1-mR_6)$ has to 
be continued to all values of $A_6$ as shown in Fig.~\ref{u1}. In the 
anomaly-free SO(10) model under discussion, we have contributions 
corresponding to a vector multiplet ${\bf 24}$, 2 hypermultiplet ${\bf 10}$s 
and 16 hypermultiplet ${\bf 5}$s in SU(5) language. Thus, the full 
correction reads 
\beq
\Delta\alpha_2^{-1}=\frac{R_5}{2R_6}\left\{3c(1-c)-3\left(\frac{c}{5}\right)
\left(1-\frac{c}{5}\right)-8\left(\frac{3}{5}c\right)\left(1-\frac{3}{5}c
\right)\right\}\,,\label{da2}
\eeq
with $c=M_VR_6$. Similar formulae for $\Delta\alpha_1^{-1}$ and $\Delta
\alpha_3^{-1}$ are easily derived using the data of Table~\ref{su5}. For 
illustration, we plot the inverse gauge coupling differences relevant to
unification in Fig.~\ref{so10}. This figure nicely illustrates the 
piecewise linear functional dependence on $A_6$ that results from a 
sum of functions of the type displayed in Fig.~\ref{u1} in an anomaly-free
model. The figure also shows that, in the specific model under 
consideration, realistic gauge unification cannot be driven by just the 
power-law effect since the ratio $\alpha_{12}/\alpha_{23}\simeq 
1.4$ is not realized for any value of $A_6$. This may be different for 
models with other matter content and corresponding Green-Schwarz anomaly
cancellation. It may also be changed if other Wilson lines or bulk 
hypermultiplet VEVs affect the mass spectrum of the model. However, since 
the main aim of the present paper is not the construction of realistic 
GUT models but rather the conceptual and technical understanding of 
power-law corrections to unification, we end our brief excursion into 
SO(10) phenomenology here. 

Finally, we emphasize that the structure of Eq.~(\ref{da2}) justifies, a 
posteriori, our assumption of an intermediate, effectively 5-dimensional
theory, i.e., the assumption $R_5\gg R_6$. Indeed, given that $c(1-c)$ and 
the other terms of this type are at most ${\cal O}(1)$, power corrections 
to inverse gauge coupling differences can only become parametrically larger 
than the familiar 4d threshold effects if $R_5/R_6\gg 1$ (cf.~\cite{bac}).

\section{Conclusion}\label{con}

We have demonstrated that power-like loop corrections to gauge coupling 
unification arising in generic supersymmetric 5d unified models are 
exactly calculable in the framework of the 5d low-energy effective 
field theory. Such power-law corrections are induced, for example, by the 
loop effects of charged bulk matter fields. They are also introduced 
by higher-dimension operators which contain the symmetry-breaking bulk 
Higgs field together with the field strength tensor. In fact, one can 
equivalently view the loop effect of bulk matter as coming from 
higher-dimension operators introduced when these fields are integrated out. 
These operators then change low-energy gauge couplings at the tree level. 

The crucial points underlying calculability are the following: On the one
hand, minimal 5d SUSY, which corresponds to ${\cal N}\!=\!2$ SUSY in 4d 
language, ensures that no corrections arise beyond the one-loop level. 
On the other hand, possible higher-dimension operators are extremely 
restricted by the combination of 5d SUSY and 5d gauge invariance. In fact,
there is only one globally analytic higher-dimension operator at the 
two-derivative level, which is the SUSY version of the Chern-Simons (CS) 
term. Knowledge of the light 5d field content and the coefficient of the 
CS term determines the low-energy gauge couplings completely. 

Realistic 5d models can arise by compactification on an interval, e.g., as 
an $S^1/Z_2$ orbifold. Since the 5d CS term induces boundary anomalies, 
a given bulk and brane field content together with the requirement that 
boundary anomalies cancel fixes the coefficient of this higher-dimension 
operator. Thus, power-like corrections to gauge coupling differences are 
completely fixed. Because of the absence of higher-loop effects or other 
higher-dimension operators, this calculability is not lost at strong 
coupling, i.e., if the gauge symmetry is broken at a scale where the 5d 
gauge theory is strongly coupled. In this case, power-law corrections 
are parametrically large and can be of the same size as the conventional 
logarithmic running from GUT scale to weak scale. In particular, we 
find that, in an SU(5) model with a single ${\bf 10}$ hypermultiplet in 
the bulk and the CS term required by anomaly cancellation, the power-law
effect is group-theoretically equivalent to MSSM running. Thus, calculable 
TeV-scale unification is possible. 

We have also considered the possibility that a 5d model arises as the 
low-energy effective theory of a 6d model compactified on an $S^1$. In this
case the 5d bulk breaking, realized in all interesting cases by the bulk
VEV of the scalar adjoint from the vector multiplet, can be traced to a
6d Wilson line wrapping the $S^1$. Such models correspond to familiar 
6d $T^2/Z_2$ constructions where the ratio of the two torus radii, $R_5$ and 
$R_6$, is taken to be large. Power-like gauge coupling corrections are 
calculable in close analogy to the 5d case and produce contributions to
differences of inverse gauge couplings of the order $\sim R_5/R_6$. 
These effective 5d theories coming from 6d are highly constrained by 6d 
anomaly cancellation and because of the absence of bulk hypermultiplet 
masses in 6d. For $d\ge 7$ the minimal SUSY corresponds to ${\cal N}\!=\!4$ 
in 4d language and no loop corrections to gauge coupling unification arise. 

Thus, we have found that large and fully calculable power-like loop 
corrections to gauge unification arise in the context of 5d and 6d 
grand unified theories. Their phenomenological relevance may be as 
striking as a lowering of the unification scale by many orders of 
magnitude or as modest as an interesting field theoretic contribution 
to the detailed GUT dynamics in a string-derived high-scale model. 
In any case, we believe that the field theoretic calculability of 
such power-like loop corrections, based on higher-dimensional SUSY, 
gauge symmetry, and anomaly cancellation, is an interesting phenomenon.

\noindent
{\bf Acknowledgements}: We would like to thank W.~Buchm\"uller, Kiwoon Choi,
D.~M.~Ghilencea, S.~Groot~Nibbelink, Hyun Min Lee, J.~Louis, 
J.~March-Russell, E.~Poppitz, and M.~Ratz for useful discussions and 
comments.

\section*{Appendix A: Cubic order of the 5d prepotential}
\setcounter{equation}{0}\renewcommand{\theequation}{A.\arabic{equation}}
The fact that the holomorphic prepotential ${\cal F}$ characterizing a 5d 
SYM theory in 4d ${\cal N}\!=\!2$ language is a polynomial of at most cubic 
order~\cite{ims} lies at the very heart of the results presented in this 
paper. We therefore include a brief discussion of this known fact. 

The lagrangian of a the 4d ${\cal N}\!=\!2$ SYM theory obtained at the zero 
mode level from a 5d theory can be written as an ${\cal N}\!=\!2$ superspace 
integral~\cite{gsw} (for a review see, e.g.,~\cite{lyk}). It is 
proportional to~\cite{seib}
\beq
\int d^2\theta\,d^2\tilde{\theta}\,{\cal F}({\cal A})\,+\,\mbox{h.c.}\,,
\eeq
where ${\cal A}$ is a Lie-algebra valued, constrained ${\cal N}\!=\!2$ 
superfield and $\theta$, $\tilde{\theta}$ are the two independent Grassmann 
coordinates. The function ${\cal F}$ is a power series in ${\cal A}$, where 
the $n$th order term
\beq
d_{a_1\cdots a_n}{\cal A}^{a_1}\cdots {\cal A}^{a_n}
\eeq
is specified by a totally symmetric invariant tensor $d_{a_1\cdots a_n}$ of
the Lie algebra. The ${\cal N}\!=\!1$ superspace expression for the 
lagrangian has been given in Eq.~(\ref{pre}).

It can now be easily shown that, if such a 4d lagrangian comes from a 
5d theory, no terms with $n>3$ can be present. Assuming the presence of a 
non-zero term with $n>3$, we first restrict our attention to a U(1) 
subgroup on whose generator the corresponding term does not vanish. The rest 
of the argument can now be given for this abelian ${\cal N}\!=\!2$ gauge 
theory. It is clear from Eq.~(\ref{prep}) that the relevant prepotential 
term induces, among others, a bosonic lagrangian term
\beq
\sim \Phi^{n-3}A_5\,\,dA\wedge dA\,,\label{aa}
\eeq
where $A$ is the 4d vector potential and $A_5$ is, for the moment, just
a 4d scalar. More explicitly, this term involves the contraction
$\epsilon^{\mu\nu\rho\sigma}\partial_\mu A_\nu\partial_\rho A_\sigma$. 
The expression in Eq.~(\ref{aa}) has to come from one of the terms of the 
original 5d component lagrangian, and this implies that the 4d 
$\epsilon$-tensor comes from a 5d $\epsilon$-tensor. Thus, the 5d component 
lagrangian contains a term
\beq
\sim \Phi^{n-3}\,A\wedge dA\wedge dA\,,\label{aaa}
\eeq
where now $A$ is the 5d vector potential. Under (abelian) gauge 
transformations, the $\Phi$-independent part transforms into a total 
derivative, so that gauge invariance of this term can not be realized unless
$n=3$. It is also easy to see that none of the other terms allowed in a 
completely generic 5d lagrangian with $\Phi$ and a vector field $A$ can 
compensate the gauge variation of Eq.~(\ref{aaa}). This completes our 
argument.

\section*{Appendix B: 6d Wilson line corrections}
\setcounter{equation}{0}\renewcommand{\theequation}{B.\arabic{equation}}
In this appendix, we present a very simple and self-contained 
derivation of the mass dependence, i.e., the Wilson line dependence, of the 
KK sum in Eq.~(\ref{wls}) (for a more detailed discussion of this and related 
calculations see, e.g.,~\cite{gng}). Recalling that each term in this sum 
comes from a 5d 1-loop integral whose only dimensionful parameter is the 
KK mass of the relevant 5d multiplet, it is immediately clear that 
performing the same calculation in $5-\epsilon$ dimensions yields
\beq
\delta\left(\frac{1}{g_{5-\epsilon}^2}\right)=-\frac{q^2}{8\pi^2}\,f(\epsilon
)\sum_{n=-\infty}^{+\infty}\,|nR_6^{-1}+m|^{1-\epsilon}\,.\label{isu}
\eeq
The function $f(\epsilon)$, the exact form of which can not be obtained 
from the above simple dimensional argument, has the limit $f(\epsilon)\to 
1$ for $\epsilon\to 0$. However, as will become obvious shortly, the 
$m$-dependent part of the infinite sum in Eq.~(\ref{isu}) is continuous in 
this limit so that the factor $f(\epsilon)$ can simply be dropped. 
Introducing the important dimensionless parameter $c=mR_6$ (we assume 
$0\!<\!c\!<\!1$ for the moment) and splitting off the $n\!=\!0$ 
contribution, the above correction can be written as 
\bea
\delta\left(\frac{1}{g_{5-\epsilon}^2}\right)&\!\!=\!\!&-\frac{q^2}{8\pi^2
R_6}\left\{c+\sum_{n=1}^\infty\left[(n+c)^{1-\epsilon}+(n-c)^{1-\epsilon}
\right]\right\}
\\
&\!\!=\!\!&-\frac{q^2}{8\pi^2R_6}\int_0^cdc'\left\{1+\sum_{n=1}^\infty
n^{-\epsilon}\left[(1+c'/n)^{-\epsilon}-(1-c'/n)^{-\epsilon}\right]\right
\}+\{\mbox{$c$-indep.}\}.\nonumber
\eea
Here we have suppressed terms ${\cal O}(\epsilon)$ whenever possible without 
affecting the final result. In the second line, an irrelevant $c$-independent 
part (the value of the sum at $c=0$) has been separated. Expanding the terms 
within the square brackets in $\epsilon$ and using the relation 
\beq
\lim_{\epsilon\to 0}\,\epsilon\sum_{n=1}^\infty n^{-1-\epsilon}=1\,,
\eeq
familiar in the context of Riemann's zeta function (see, e.g.,~\cite{int}),
one finds the result given in Eq.~(\ref{6du1}). 

Before closing, we would like to comment on the regularization independence 
of the above result. In fact, returning to the level of actual loop 
integrations, the calculation can be performed without ever introducing a 
regularization. It is clear that the desired $A_6$ dependence of the 
gauge coupling can be extracted from the $c$ dependence of a sum of 5d 
one-loop integrals, conveniently written as the integral of a sum, of the 
form
\bea
I(c)&\!\!=\!\!&\int\frac{d^5k}{(2\pi)^5}\sum_{n=-\infty}^{+\infty}\frac{1}
{\left[k^2+(n+c)^2\right]^2}=\int\frac{d^5k}{(2\pi)^5}\sum_{n=-\infty}^{+
\infty}\left\{-\frac{\partial}{\partial k^2}\left(\frac{1}{k^2+(n+c)^2}
\right)\right\}\label{last}
\\
&\!\!=\!\!&\int\frac{d^5k}{(2\pi)^5}\left\{-\frac{
\partial}{\partial|k|^2}\left(\frac{1}{|k|}\cdot\frac{\pi\,\sinh(2\pi |k|)}
{\cosh(2\pi |k|)-\cos(2\pi c)}\right)\right\}=-\frac{c(1-c)}{16\pi^2}+
\{\mbox{$c$-indep.}\}\,.\nonumber
\eea
Here we have focussed on the simplest scalar integral appearing in the 
detailed calculation, rescaled the 5-momentum according to $k\to k/R_6$,
and suppressed an overall $A_6$-independent factor. Thus, all we need is
$I'(c)$, which is finite simply because the first derivative with respect to
$c$ of the integrand in Eq.~(\ref{last}) falls exponentially for $|k|\to 
\infty$. We have used {\it Mathematica}~\cite{wol} for evaluating the sum 
and the integral. Note also that dropping all higher KK modes (i.e., 
restricting the sum to $n=0$) corresponds to the replacement $c(1-c)\to c$ in 
the final answer. 

Performing the sum before the (in general divergent) 5d loop integration is
crucial because in this way we are sure to respect the non-locality of the
Wilson line effect in the 6d theory. This non-locality is the reason for
finiteness. Regularization is just useful for finding the explicit result
in a somewhat simpler way, not necessary at the conceptual level. Of course, 
it has to respect the non-local structure of the Wilson line wrapping the 
$S^1$. Clearly, dimensional regularization in the 5 non-compact dimensions 
satisfies this requirement. We could have reduced all relevant loop 
corrections to the form of Eq.~(\ref{last}) to arrive at our result. 
However, in this paper we have emphasized the dimensional regularization
approach as the simplest way to obtain the answer directly from the known 
5d formula.

\end{document}